# 3D NUMERICAL RELATIVITY AT NCSA[*]

*PETER ANNINOS*[1], *KAREN CAMARDA*[1], *JOAN MASSÓ*[1,2], *EDWARD SEIDEL*[1],
*WAI-MO SUEN*[3], *MALCOLM TOBIAS*[3], *JOHN TOWNS*[1]

[1] *National Center for Supercomputing Applications*
*605 E. Springfield Ave., Champaign, Illinois 61820*
[2] *Departament de Física, Universitat de les Illes Balears,*
*E-07071 Palma de Mallorca, Spain*
[3] *McDonnell Center for the Space Sciences, Department of Physics*
*Washington University, St. Louis, Missouri, 63130*

ABSTRACT

Numerical relativity is finally coming of age with the development of massively parallel computers. 3D problems, which were completely intractable several years ago due to limited computer power, can now be performed with grid sizes of about $200^3$. We report on several new codes developed for solving the full 3D Einstein equations, and present results of using them to evolve black holes and gravitational waves.

## 1. Introduction

The Einstein equations for the gravitational field were first published near the beginning of this century. Unfortunately, the equations are a set of 10 nonlinear, coupled, hyperbolic-elliptic partial differential equations that are not amenable to analytic study except in highly idealized cases involving symmetries or approximations. Similarly, they also pose a major challenge to numerical relativity, as the equations are complex and the variables so many that until quite recently accurate numerical treatment has been restricted to 1D and 2D studies. For these reasons, in spite of many years of study, the solution space of the complete set of the Einstein equations is very much unknown.

A new generation of high performance, massively parallel computers, such as the NCSA Connection Machine CM-5 and others like it, promises to revolutionize computational physics, and numerical relativity, during this decade. Based on scalable, parallel technology, machines are now available that can achieve tens of GFLOPs on actual application codes. This is about 100 times the speed attainable just a few years ago on a Cray Y-MP, due to both hardware and algorithm improvements. Moreover, the memory of the machines is increasing along with the speed; large machines now provide tens of Gbytes of memory, dozens of times that of a Y-MP, allowing much larger problems to be solved. To take advantage of such machines for numerical relativity, physicists and computer scientists beginning to work closely together to develop new algorithms to solve the Einstein equations on parallel computers.

---

## 2. 3D Codes

We have developed two 3D codes in cartesian coordinates to study the most general possible spacetimes. One of these codes uses the standard ADM formalism with arbitrary shift and slicing conditions, and the other uses the 3D harmonic formulation of Bona and Massó[1] that puts the equations in an explicitly hyperbolic, first order, flux-conservative form. This formulation allows advanced numerical techniques for such systems to be applied to the Einstein equations for the first time, and has recently being extended to a wide class of slicing conditions.[2] These codes are now being used for 3D studies of black holes and gravitational waves, as we report here.

These codes for solving the Einstein equations are well suited to parallel machines. The harmonic code is one of the best performers of any application running on the NCSA 512 node CM-5, presently achieving nearly 20GFlops. It is also one of the fastest application codes running on the Pittsburgh 16 processor Cray C-90, where it achieves 8 GFlops.

## 3. 3D Black Holes

In this section we present results for evolving 3D black hole spacetimes.[3] These calculations are very difficult because of the sharp peaks of the metric functions that develop when evolving black holes if one uses a singularity avoiding time slicing. We identify a number of difficulties in evolving 3D black holes and suggest approaches to overcome them. We show how special treatment of the conformal factor can lead to more accurate evolution, and discuss a number of techniques we developed to handle black hole spacetimes in the absence of symmetries.

We have tested many different slicing conditions for 3D black hole evolution, including geodesic, maximal, and various algebraic conditions on the lapse. Geodesic slicing is used as code test, since the solution is known analytically. The code is able to evolve the black hole with errors of less than a few percent until the time the slices hit the singularity. With current resolutions, limited by computer memory sizes, we show that with certain lapse conditions we can evolve the black hole to about $t = 50M$, where $M$ is the black hole mass. Comparisons are made with results obtained by evolving spherical initial black hole data sets with a 1D spherically symmetric code.

We have also implemented an apparent horizon boundary condition which can be used to prevent the development of large gradients in the metric functions that result from singularity avoiding time slicings. We compute the mass of the apparent horizon in these spacetimes, and find that in many cases it can be conserved to within about 5% throughout the evolution with our techniques and current resolution.

In Fig. 1 we show a slice $z = 0$ of the conformal radial metric function $g_{rr}$ at $t = 50M$ for a Schwarzschild black hole placed in the center of a 3D grid, evolved with an algebraic slicing.

## 4. 3D Gravitational Waves

We have also been using the same 3D codes to the problems of pure gravitational waves.[4] There are theoretical, observational, and technical reasons to study pure wave spacetimes. A pure wave system will allow us to probe the theory of general relativity itself in the nonlinear regime. This important area of research is for the most part uncharted territory. Fundamental questions about 3D gravitational waves, involving for example gravitational geons or the formation of singularities, are unknown. Previous analytic and numerical work on pure gravitational wave spacetimes, done in one or two spatial dimensions, has led to many interesting discoveries, such as the formation of singularities from colliding plane waves or the existence of critical behavior in black hole formation in axisymmetric spacetimes. These discoveries raise interesting questions about what will happen in the generic 3D case.

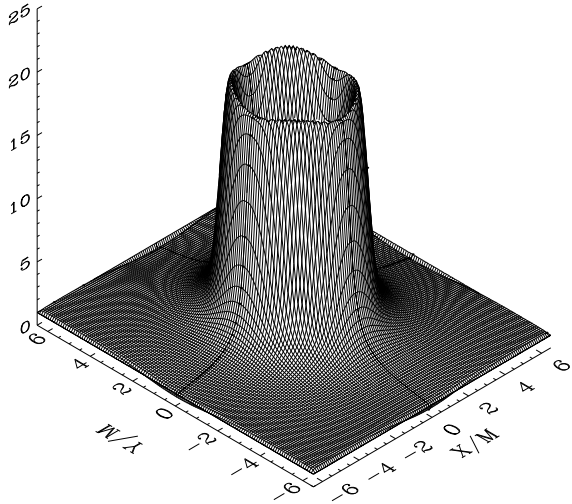

Figure 1: Schwarzschild black hole component $g_{rr}$.

We have developed both fully nonlinear and linearized versions of both of our codes, so that we can compare differences and sort out nonlinear physics and code effects. As further tools to analyze results and to confirm their accuracy, we used Fourier analysis to see how the codes propagate various frequency components, and we monitored the constraints, which should remain small throughout the evolution. At present, our codes have passed all testbeds devised, and we are in the beginning phase of using the codes to investigate new physics.

In Fig. 2 we show an isosurface of the 3d metric function $g_{xx}$ for an initially linear 3D gravitational wave evolved with our H code. The image is a snapshot from a virtual reality environment (CAVE) application to analyze 3D spacetimes that we are developing[5] in collaboration with the Electronic Visualization Laboratory at Chicago.

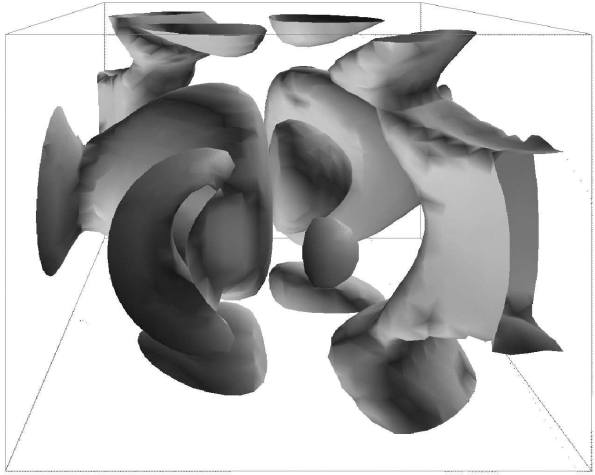

Figure 2: CAVE snapshot of gravitational wave component $g_{xx}$.

## 5. Conclusions

With these codes described above, and the rapid development of massively parallel computers, we expect that general 3D gravitational systems, with their full nonlinearities, can be studied numerically within the next few years. This opens the possibility of many interesting discoveries in physics of systems ranging from black holes to the universe as a whole.

### Acknowledgements

We acknowledge the support of NCSA and NSF grants PHY94-07882, 94-04788 and PHY/ASC93-18152 (arpa supplemented). J.M. also acknowledges a Fellowship (P.F.P.I.) from Ministerio de Educación y Ciencia of Spain.

### References

1. C. Bona and J. Massó, Phys. Rev. Lett. **68**, 1097 (1992).
2. C. Bona, J. Massó, J. Stela, and E. Seidel, "A Class of Hyperbolic Gauge Conditions", in these proceedings.
3. P. Anninos, K. Camarda, J. Massó, E. Seidel, W.-M. Suen, J. Towns, Phys. Rev. D., submitted (1994).
4. P. Anninos, J. Massó, E. Seidel, W.-M. Suen, M. Tobias, Phys. Rev. D., in preparation.
5. J. Goldman, T. Roy, J. Massó, E. Seidel, "Spacetime Splashes: Catching the Wave in Einstein Equations", VROOM demonstration at SIGGRAPH 94, Florida.